\documentclass[aps,prb,reprint,amsmath,amssymb,showpacs,superscriptaddress]{revtex4-1}
\usepackage{amsmath}
\usepackage{amsfonts}
\usepackage{amssymb}
\usepackage{mathtools}
\usepackage{pict2e}
\usepackage{hyperref}
\usepackage{graphicx,xcolor}
\usepackage{bbding}
\usepackage{bm}
\usepackage{multirow}
\usepackage{grffile}

\newcommand{\bs}[1]{\boldsymbol{#1}}
\mathchardef\sGamma="7100
\mathchardef\sOmega="710A
\def\nd{^{\vphantom{\dagger}}}
\def\BM{{\bs{M}}}

\def\BOh{{\bs{{\sOmega}}}}
\def\Br{{\bs{r}}}
\def\Bnabla{{\bs{\nabla}}}
\def\pz{\partial}
\def\lab{{\bar\lambda}}
\def\wb{{\bar w}}

\def\CE{{\mathcal{E}}}

\def\CAE{{\mathcal{A}_{\rm E}}}
\def\CAF{{\mathcal{A}^{\rm eff}_{\rm E}}}
\def\CD{{\mathcal{D}}}
\def\half{\frac{1}{2}}
\def\ns{^{\vphantom *}}
\def\xhat{{\hat{\bs{x}}}}
\def\yhat{{\hat{\bs{y}}}}
\def\zhat{{\hat{\bs{z}}}}
\def\Hbar{{\bar H}}

\def\sexpect#1#2#3{{\langle \, #1 \, | \,  #2  \, | \, #3 \, \rangle}}

\begin{document}

\title{Quantum Nucleation of Skyrmions in Magnetic Films by Inhomogeneous Fields}

\author{Sebastian A. Diaz}
\email[]{sdiaz@physics.ucsd.edu}
\affiliation{Department of Physics, University of California, San Diego, La Jolla, California 92093, USA}

\author{Daniel P. Arovas}
\email[]{darovas@ucsd.edu}
\affiliation{Department of Physics, University of California, San Diego, La Jolla, California 92093, USA}

\date{\today}

\begin{abstract}
Recent experiments have reported on controlled nucleation of individual skyrmions in chiral magnets.  Here we show that in magnetic ultra-thin films
with interfacial Dzyaloshinskii-Moriya interaction, single skyrmions of different radii can be nucleated by creating a local distortion in the magnetic field.
In our study, we have considered zero temperature quantum nucleation of a single skyrmion from a ferromagnetic phase. The physical scenario we model
is one where a uniform field stabilizes the ferromagnet, and an opposing local magnetic field over a circular spot, generated by the tip of a local probe,
drives the skyrmion nucleation. Using spin path integrals and a collective coordinate approximation, the tunneling rate from the ferromagnetic to the single skyrmion state is computed as a function of the tip's magnetic field and the circular spot radius. Suitable parameters for the observation of the quantum nucleation of single skyrmions are identified.
\end{abstract}

%\pacs{}

\maketitle

%%%%%%%%%%%%%%%%%%%%%%%%%%%%%%%%%%%%%%%
%%%%%%%%%%%%%%%%%%%%%%%%%%%%%%%%%%%%%%%

\section{Introduction}

%What are magnetic skyrmions? 
Magnetic skyrmions (also known as ``baby skyrmions") are two-dimensional topological configurations in which the direction $\BOh(\Br)$ of
magnetization field $\BM(\Br)$ wraps around the unit sphere.  More precisely, $\BM(\Br)$ supports a single skyrmion/antiskyrmion when its
integer-valued topological charge, or Pontrjagin index, 
\begin{equation}
Q = \frac{1}{4\pi}\int\!d^2\!r \>\BOh\cdot{\pz\BOh\over\pz x}\times{\pz\BOh\over\pz y}\quad,
\end{equation}
is equal to $\pm 1$, with $\BOh(\Br)=\BM(\Br)/|\BM(\Br)|$.  Skyrmions are stable against smooth variations of the magnetization since $Q$ cannot jump
continuously from one integer value to another.  Because skyrmions are localized in space and are topologically stable, they behave as particles.  

%Experimental observation, novel properties and applications.
In certain chiral magnets, the competition between local exchange, Dzyaloshinskii-Moriya interactions (DMI), and external magnetic field stabilizes a
skyrmion crystal phase, in which $Q$ is thermodynamically large and given by the number of magnetic unit cells of the structure. Such configurations
were first observed in bulk MnSi by neutron scattering \cite{Muhlbauer2009}, and real space observation of skyrmions was achieved in other chiral
magnets using Lorentz transmission electron microscopy (TEM) \cite{Yu2010}. Owing to their topological nature,  systems supporting skyrmions
exhibit novel properties such as emergent magnetic monopoles \cite{Milde2013} and electromagnetic fields \cite{Schulz2012}, as well as the
topological \cite{Neubauer2009} and skyrmion \cite{Zang2011} Hall effects. Their microscopic size, topologically-protected stability and effective
coupling to electric currents make skyrmions attractive for applications. Their small depinning current densities \cite{Jonietz2010}, some six orders of
magnitude smaller than those needed for domain walls, and their ability to move around obstacles \cite{Fert2013,Iwasaki2013b} make them promising information carriers in magnetic storage and logic devices, where their motion can be controlled by currents or electric fields.

%Skyrmion nucleation motivation.
Recent experiments have demonstrated the ability to nucleate skyrmions in a controlled fashion. In thin magnetic films, skyrmions can be nucleated at
the sample edge by spin-polarized electric currents \cite{Yu2012}. Using spin-polarized tunneling from the tip of a scanning tunneling microscope (STM),
skyrmions have been written and erased from ultra-thin magnetic films \cite{Romming2013}. Another experiment uses an in-plane electric current to force
magnetic stripe domains through a geometrical constriction to nucleate skyrmions in a process resembling soap bubble blowing \cite{Jiang2015}.
Finally, time-dependent magnetic fields generated by sending current pulses down a microcoil were used to nucleate skyrmions in magnetic disks
\cite{Woo2016}.

%Relevance of skyrmion nucleation
Topological charge is created during the skyrmion nucleation process.  Although smooth deformations of the magnetization field cannot make $Q$ jump
from one integer value to another in the continuum, this restriction does not apply to spins on a lattice \cite{PhysRevLett.61.1029}.  Here, we shall
derive the instanton paths corresponding to skyrmion nucleation in a lattice model, using a collective coordinate approach. From an application-oriented
perspective, understanding and controlling the nucleation of skyrmions should be of importance in exploiting these textures as information carriers.
Thus far, experimental and theoretical studies have relied on electric currents, time-dependent magnetic fields, and local heating to provide the energy injection necessary to overcome the energy barrier preventing the system from reaching the magnetic skyrmion texture. However, even in the absence of thermal fluctuations or external perturbations, the system has a nonzero probability to escape from a metastable state by quantum tunneling through the energy barrier. 

\begin{figure}[h!]
\centering
\includegraphics[width=0.65\columnwidth]{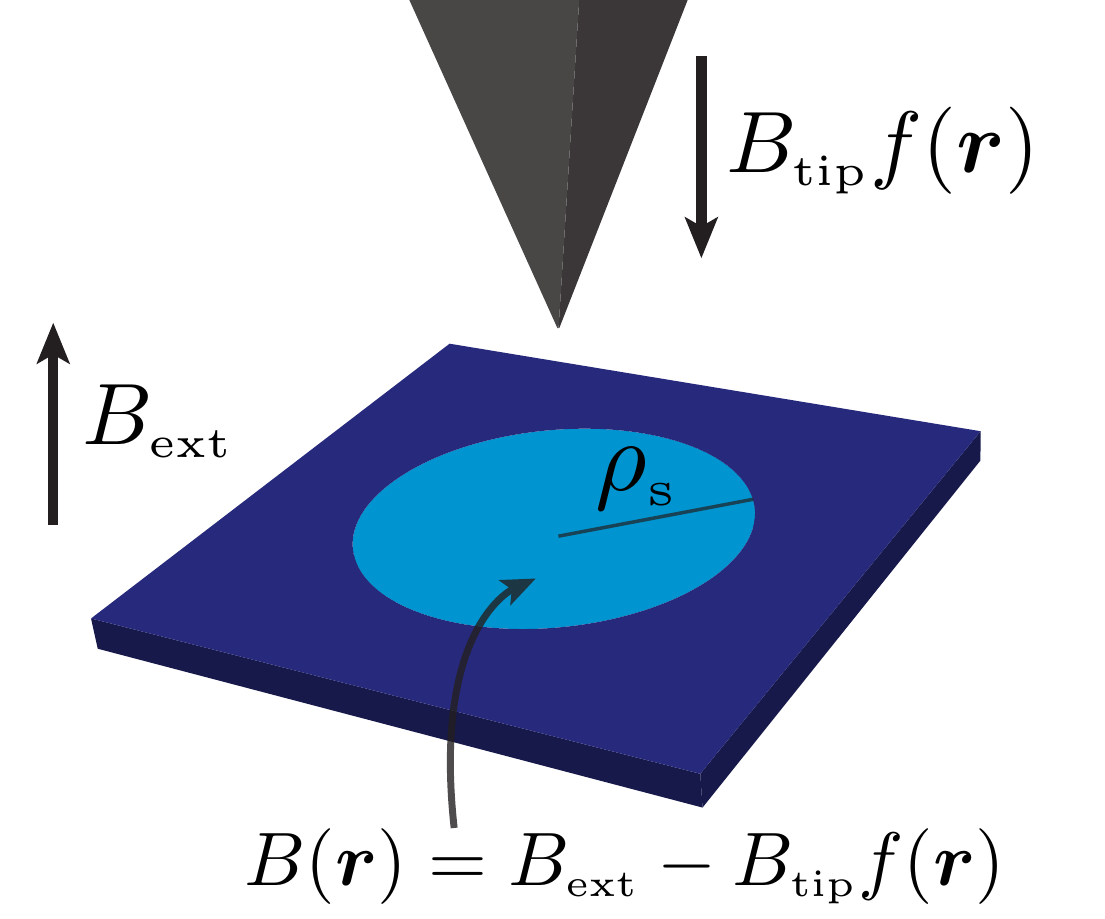}
\caption{{\bf Setup Schematic.} A uniform field, $B_{{\rm ext}}$, and an opposing local magnetic field, generated by the tip of a local probe, $B_{{\rm tip}}f(\Br)$, are simultaneously applied to the sample. Thus the net local magnetic field is $B(\Br)=B_{\rm ext} - B_{\rm tip}f(\Br)$. The dimensionless form factor $f(\Br)$ is such that within the circular spot with radius $\rho_{\rm s}$, $B(\Br) \simeq B_{\rm ext} - B_{\rm tip}$, while $B(\Br) \simeq B_{\rm ext}$ outside.}
\label{fig:Setup}
\end{figure}

%Paper structure and summary of our findings. 
Here we model skyrmion nucleation in magnetic ultra-thin films with interfacial Dzyaloshinskii-Moriya interaction (DMI).  In the continuum limit, the energy density
is\cite{Yu2010} ${\cal E}(\Br)=\half J (\Bnabla\BM)^2 + D\big[M^z\Bnabla\cdot\BM - (\BM\cdot\Bnabla)M^z\big] - B M^z$, with $B(\Br)$ initially uniform.  At zero temperature, the low $B$ phase is a helical structure (H) and the high $B$ phase is a uniformly magnetized ferromagnet (FM).  Interpolating these phases
is the skyrmion crystal (SkX).  We consider the effect of a local reduction in $B$ close to the SkX-FM boundary, as described in section II below.
If the local field is sufficiently reduced over a sufficiently large spot, we find that the lowest energy state is one accommodating a single skyrmion (SSk).
In section III, we introduce a lattice-based version of this model in which the individual spins are endowed with quantum dynamics.  Then using spin path
integrals and a collective coordinate approximation, we compute the rate at which single skyrmions are nucleated out of the metastable FM configuration.  Section IV contains the results and discussion of the tunneling rate calculation, followed by conclusions. 

%%%%%%%%%%%%%%%%%%%%%%%%%%%%%%%%%%%%%%%
%%%%%%%%%%%%%%%%%%%%%%%%%%%%%%%%%%%%%%%
\section{Setup and System Preparation}

%Setup description.
Magnetic skyrmion lattices have been observed via neutron scattering in thin films of chiral magnets, such as MnSi and Fe$_{1-x}$Co$_x$Si, and via spin-polarized
STM in monolayer Fe films on Ir(111) surfaces.  In the latter case, the skyrmion size is of atomic dimensions.  Since the nucleation rate is expected to scale 
exponentially with the number of spins involved in the tunneling process, here we are interested in fairly compact skyrmions.  The experimental setup we envision
is depicted in Fig. \ref{fig:Setup}.   A uniform magnetic field, $B_{\rm ext}$, is applied perpendicular to the film ($\zhat$) to control its global magnetic phase. 
A local probe, such as a magnetized STM tip, is then invoked to provide an additional opposing field $-B_{\rm tip}f(\Br)$, also oriented along $\zhat$ in our model,
where $f(\Br)$ is a dimensionless form factor describing the field distribution.  Thus, the net local magnetic field is $B(\Br)=B_{\rm ext} - B_{\rm tip}f(\Br)$.  We assume
the field from the tip is azimuthally symmetric, {\it i.e.\/} $f=f(\rho)$ with $\rho=|\Br|$, and that $f(\rho)\simeq 1$ inside a disk (`spot') of radius $\rho_{\rm s}$, dropping
to zero rapidly for $\rho>\rho_{\rm s}$.  Specifically, we take
\begin{equation}
f(\rho)=\Theta(\rho_{\rm s}-\rho)+e^{-(\rho-\rho_{\rm s})^2/\sigma^2}\Theta(\rho-\rho_{\rm s})\ ,
\end{equation}
where $\Theta(x)$ is the step function. The quantities $B_{\rm ext}$, $B_{\rm tip}$, $\rho_{\rm s}$, and $\sigma$ are all adjustable parameters within our model.

%System preparation.
To prepare the system for the quantum nucleation of a SSk, a uniform magnetic field is first applied to bring the sample to the FM state. Then, the local probe field is turned on. Varying the probe field strength and the spot radius, it can be shown that a SSk can become energetically more favorable than the FM. Thus, under these conditions, the system is still ferromagnetically ordered, but now in a metastable state. At zero or low enough temperature, thermal fluctuations cannot overcome the energy barrier separating the FM and SSk states. However, quantum tunneling renders the FM state unstable, hence there is a nonzero probability to decay to the SSk state.  

%%%%%%%%%%%%%%%%%%%%%%%%%%%%%%%%%%%%%%%
%%%%%%%%%%%%%%%%%%%%%%%%%%%%%%%%%%%%%%%
\section{Theoretical Model}

%%%%%%%%%%%%%%%%%%%%%%%%%%%%%%%%%%%%%%%%
\subsection{Discretization and quantum mechanical action}
Since the Pontrjagin index cannot change under a smooth deformation of the field $\BOh(\Br)$, in order to accommodate quantum tunneling between topological
sectors in our model, we must account for the underlying discrete lattice on which the spins are situated\cite{PhysRevLett.61.1029}.  We discretize on a square lattice of lattice constant $a$, and
henceforth we measure all lengths in units of $a$. We assume the magnitude $|\BM(\Br)|$ of the local magnetization is fixed at $M_0$, which is set by a consideration of
interaction effects\cite{Muhlbauer2009}.  After introducing a uniaxial anisotropy term $- K (M^z)^2$ and identifying an energy scale $E_0=t J M_0^2$, with $t$ the thickness of the film, the discretized dimensionless
energy, $\Hbar=E/E_0$\,, with $E=t\!\int \!d^2\!r\> \CE(\Br)$\,, takes the form
\begin{align}\label{eq:ClassicalH}
\Hbar & = - \sum_{\Br}\! \Big\{ \BOh_{\Br}\!\cdot\big(\BOh_{\Br + \xhat} + \BOh_{\Br + \yhat}\big) + \alpha \Big[ \xhat\cdot\BOh_{\Br}\times\BOh_{\Br + \yhat} \nonumber \\
&\hskip 0.4in - \yhat\cdot\BOh_{\Br}\times\BOh_{\Br + \xhat}  \Big] + \kappa\, (\sOmega^z_{\Br})^2 + b_\Br\, \sOmega^z_\Br\Big\}\ ,
\end{align}
where $\alpha=aD/J$ is a ratio of the lattice constant to the length scale $R_0=J/D$ set by the competition between DMI and exchange terms, $\kappa=Ka^2/J$, and $b_\Br=B(\Br)/B_0$ with $B_0 = J M_0/a^2$ a magnetic field scale.  

To endow our model with quantum dynamics, we extend each $\BOh_\Br$ to a function of imaginary time, and write the quantum action,
\begin{equation}
\!\!\CAE\big[\{\BOh_\Br(\tau)\}\big]=\!\!\int\limits _0^{\beta E_0}\!\!\!d\tau \Bigg\{ iS\sum_\Br{d\omega_\Br(\tau)\over d\tau} + \Hbar\big[\BOh_\Br(\tau)\big]\Bigg\}\ ,
\label{action}
\end{equation}
where $\beta$ is the inverse temperature, $d\omega_\Br(\tau)/d\tau$ is the rate at which solid angle is swept out by the evolution of $\BOh_\Br(\tau)$, and $S$ is the
quantized spin value.  Note that we have rescaled imaginary time by units of $\hbar/E_0$ in order to render it dimensionless.

\begin{figure}[h!]
\centering
\includegraphics[width=\columnwidth]{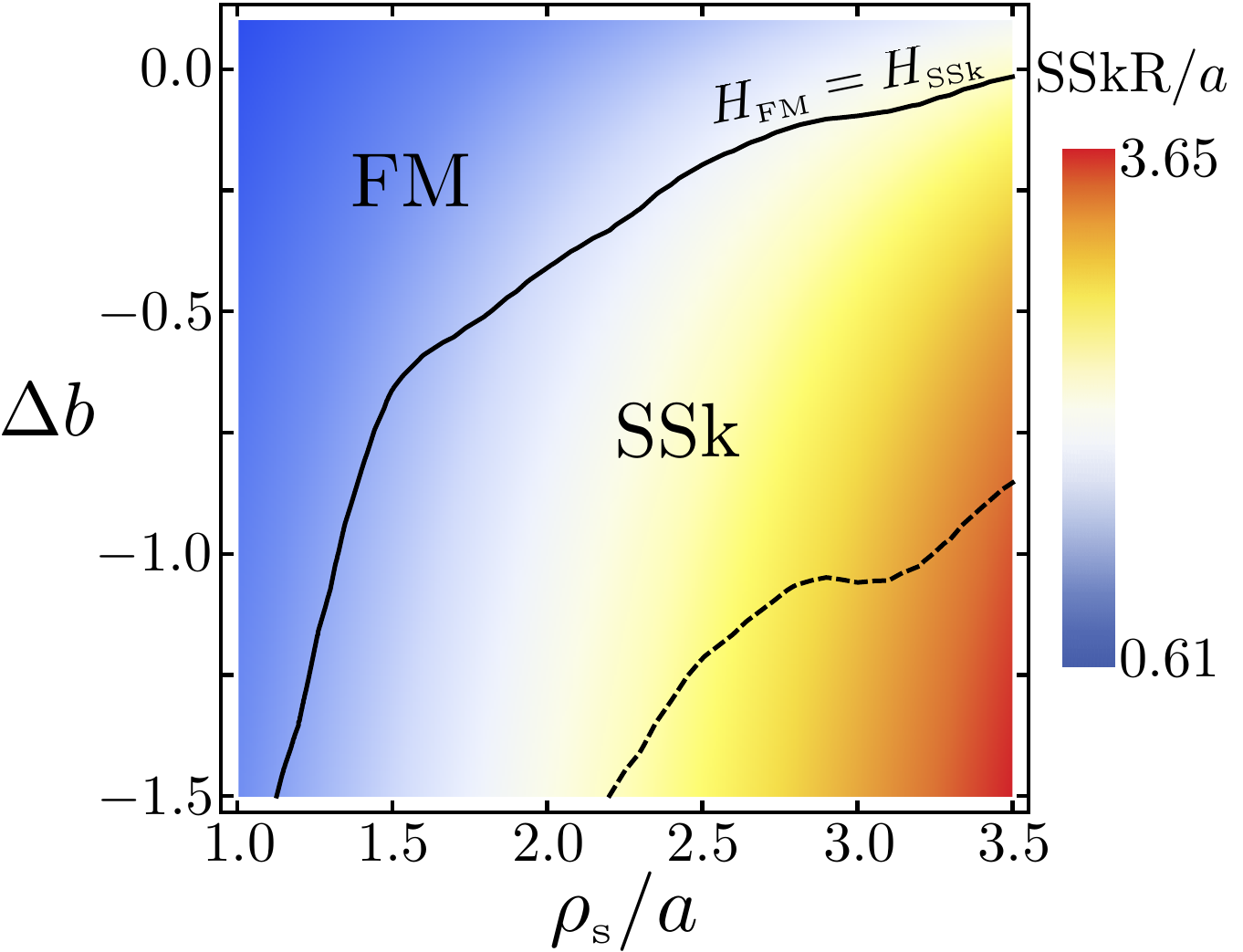}
\caption{{\bf Phase Diagram and SSk Radius}. Zero temperature phase diagram of the model Hamiltonian from Eqn. \eqref{eq:ClassicalH}, with
$\{ \alpha, \kappa, b_\infty, \sigma/a \}= \{0.5, 0.1, 3.0, 0.5\}$, for a system comprising 20$\times$20 spins on a square lattice with periodic boundary conditions.
Parameters are such that the FM and the SSk are the states with lowest energy. Above/Below the dashed, black curve within the SSk region, the FM
state is metastable/unstable. The overlaid density plot corresponds to the SSk radius, SSkR, in units of the lattice constant $a$.}
\label{fig:PhDiag}
\end{figure}

%%%%%%%%%%%%%%%%%%%%%%%%%%%%%%%%%%%%%%%%
\subsection{Phase Diagram}

%Need to determine region in parameter space where the FM becomes a metastable state with the single skyrmion as the lowest energy state.
In order to study quantum tunneling from the FM to the SSk we first identify the values of the model parameters such that the SSk is the lowest energy state
and the FM is a metastable state. The energies of these two states were computed numerically, using the Hamiltonian $\Hbar\big(\{\BOh_\Br\}\big)$ in Eqn. \eqref{eq:ClassicalH}.
For the FM, the spin at every lattice site points along the $\zhat$ direction.  The SSk spin configuration was first obtained from solving the problem in the
continuum limit (see Appendix \ref{appendix}) and then mapped to the square lattice. Computing the FM and SSk energies as a function of the model parameters, we were able to determine the appropriate region in parameter
space suitable for our tunneling study.  Fig. \ref{fig:PhDiag} shows the zero temperature phase diagram for 20$\times$20 spins (periodic boundary conditions), with $\{ \alpha, \kappa, b_\infty, \sigma/a \} = \{0.5, 0.1, 3.0, 0.5\}$, $b_\infty = B_{{\rm ext}}/B_0$ and $\Delta b = (B_{{\rm ext}} - B_{{\rm tip}})/B_0$, indicating regions where the FM and SSk configurations are the ground states. As the spot size $\rho_{\rm s}$ increases, it becomes energetically favorable for more
skyrmions to form, and as $\rho_{\rm s}\to\infty$ one of course obtains the SkX phase over a range of values of $B$.  Another possibility is for the spot to
support a helical state droplet.  Accordingly, we have focused on regions in parameter space where the FM and SSk states are the only relevant ones.

As expected, we find the SSk state is the ground state over a region where $\rho_{\rm s}$ is moderate and $\Delta b$ is sufficiently small or negative, below the solid black line as shown in Fig. \ref{fig:PhDiag}. In the region between the solid and dashed curves, the FM state is metastable. When the dimensionless profile $f(\Br)$ is sharp, the phase boundaries become jagged, since the number of lattice sites within the spot changes discontinuously with the spot radius. The phase diagram is overlaid with a density plot of the SSk radius, defined as the distance from the skyrmion center to the point where the spins lie in the plane of the magnetic film. This shows that the size of the quantum nucleated SSk can be adjusted by tuning the localized magnetic field parameters. For the model parameters chosen, the skyrmions extend over just a few lattice spacings.
 
%%%%%%%%%%%%%%%%%%%%%%%%%%%%%%%%%%%%%%%%
\subsection{Tunneling Rate and Collective Coordinates}

According to a classical description, once in the FM metastable state, in the absence of thermal fluctuations, it is impossible for the system to overcome the energy
barrier that separates it from the SSk state. However, within a quantum mechanical description, the FM state is rendered unstable due to quantum tunneling.
This process is analyzed by looking at the survival probability amplitude that the system remains in the FM state after a long time $T$, {\it i.e.\/}
$\sexpect{{\rm FM}}{e^{-iHT/\hbar}}{{\rm FM}}$, where $H$ is the quantum Hamiltonian.  The magnitude of the tunneling amplitude decays for large $T$ as
$\exp(-\sGamma T/2)$, where $\sGamma$ is the inverse lifetime of the metastable FM state.We will calculate $\sGamma$ using path integrals and the standard
technique of instantons \cite{Coleman1985}. To that end we first write the above probability amplitude as the following multi-spin coherent state path integral in
Euclidean time,
\begin{equation}\label{eq:EcldSurvAmplt}
\sexpect{{\rm FM}}{e^{-\beta H}}{{\rm FM}} = \int\! \CD\BOh(\tau)\,e^{-\CAE[\BOh(\tau)]}\quad,
\end{equation}
where $\hbar\beta = iT$, $\BOh = \{ \BOh_{\Br} \}$, and the dimensionless Euclidean action $\CAE\big[\BOh(\tau)\big]$ is given in Eqn. \eqref{action}.
Na{\"\i}vely, the boundary conditions on the path integral are $\BOh(0)=\BOh(\beta E_0)=\BOh_{\rm FM}$, but as the action is linear in time derivatives, supplying
initial and final conditions is problematic.  However, for long time bounce paths as we shall consider below, the instantons come very close to satisfying these conditions.
A careful discussion of boundary conditions on the spin path integral is provided in the work of Braun and Garg in Ref. \onlinecite{BraunGarg2007}.

Rather than expressing the path integral in terms of the unit vectors $\BOh_\Br$, we find it useful to instead use their stereographic projections $w_\Br=v_\Br/u_\Br$,
where $u_\Br=\cos(\theta_\Br/2)$ and $v_\Br=\sin(\theta_\Br/2)\exp(i\phi_\Br)$ are spinor coordinates for the spin at site $\Br$.  Let $\big(u_\Br^{\rm Sk},v_\Br^{\rm Sk}\big)$
be the spinor coordinates corresponding to a static single skyrmion configuration $\BOh_\Br^{\rm Sk}$ which extremizes the energy function $\Hbar$ in Eqn.
\eqref{eq:ClassicalH}.  Rather than attempting to solve for the full instanton, we adopt here a simplifying collective coordinate description, parametrized by a complex scalar
$\lambda(\tau)$, and we write
\begin{equation}
w_\Br(\tau)={\lambda(\tau)\,v_\Br^{\rm Sk}\over 1 + \lambda(\tau)\,u_\Br^{\rm Sk}} \quad,\quad
{\bar w}_\Br(\tau)={\lab(\tau)\,{\bar v}_\Br^{\rm Sk}\over 1 + \lab(\tau)\,{\bar u}_\Br^{\rm Sk}} \ .
\end{equation}
For $\lambda\to 0$, we have that $w_\Br$ describes a FM configuration, while for $\lambda\to\infty$ $w_\Br$ corresponds to a single skyrmion.  For $\lambda=\pm 1$
we encounter a singularity in the continuum limit, where the Pontrjagin index changes discontinuously.  On the lattice, however, the singularity is avoided by discretizing
in such a way that no lattice point lies at the origin.  The central plaquette then lies at the spatial center of the space-time hedgehog defect responsible for the change in topological index \cite{PhysRevLett.61.1029}.  

Our collective coordinate description provides us with a rather simple description of the topology change via quantum nucleation.
A more complete description of our instanton, accounting for the full dynamics of the spin field, such as in
Ref. \onlinecite{PhysRevLett.79.5054}, will be discussed in a future publication.

\begin{figure}[!t]
\centering
\includegraphics[width=0.98\columnwidth]{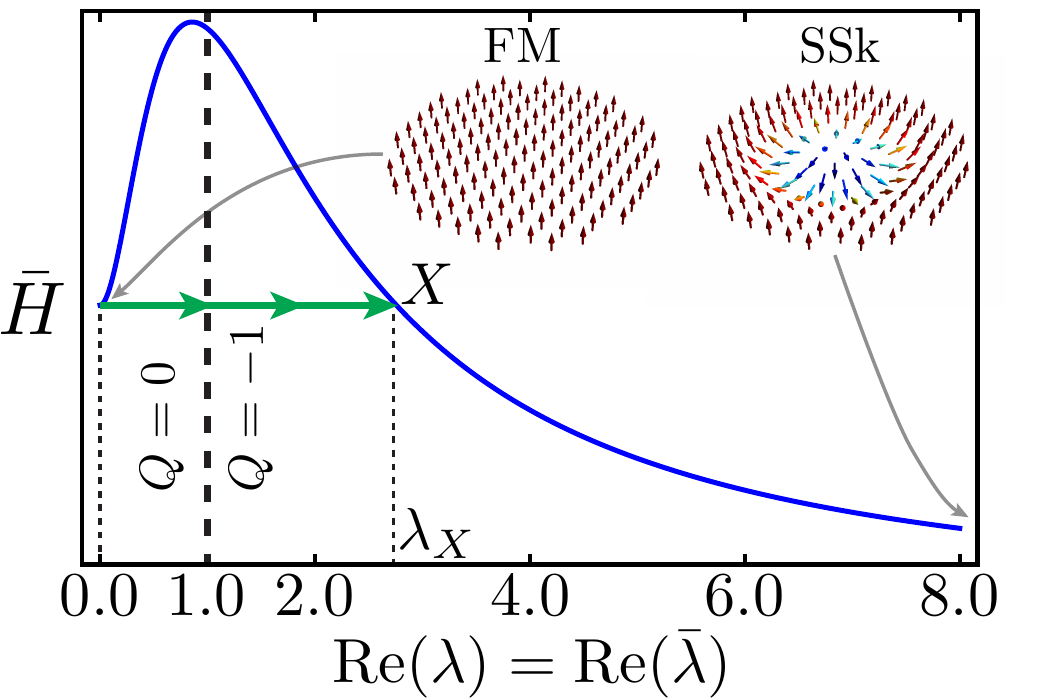}
\caption{{\bf Quantum Tunneling Process}. Energy landscape as a function of the collective coordinates along the line $\text{Re}(\lambda) = \text{Re}(\bar{\lambda})$, with $\text{Im}(\lambda) = 0 = \text{Im}(\bar{\lambda})$, for the model parameters $\{ \alpha, \kappa, b_\infty, \sigma/a, \rho_{\rm s}/a, \Delta b \}= \{0.5, 0.1, 3.0, 0.5,1.5, -0.85\}$. The classically metastable FM state, at $\lambda = 0 = \bar{\lambda}$, is rendered unstable due to quantum tunneling. The system can tunnel through the energy barrier to the $X$ state, at $\lambda_X$, and then reach the SSk state, at $\lambda, \bar{\lambda}\to\infty$, by classical evolution. The topological charge of the $X$ state is $-1$ or $0$, if $\lambda_X > 1$ or $\lambda_X < 1$, respectively.}
\label{fig:QT}
\end{figure}

By investing the interpolation parameters $\{\lambda,\lab\}$ with time dependence, the evolution of the set $\big\{w_\Br(\tau),\wb_\Br(\tau)\big\}$ now depends on the evolution of these
new collective coordinates.  Our approach here parallels that taken in Ref. \onlinecite{PhysRevA.59.1461}, where a collective coordinate path integral approach was applied
to analyze metastable Bose-Einstein condensates.  The situation is depicted in Fig. \ref{fig:QT}.  As a function of real $\lambda$, the energy $\Hbar(\lambda)$ has
a local minimum at the FM state $\lambda=0$, and a global minimum at the skyrmion state $\lambda=\infty$ (blue curve).  In the quantum tunneling process, the collective
coordinate moves under the barrier, emerging at the point marked $X$ in the figure, at which point it may `roll downhill' toward $\lambda=\infty$.  At the conclusion of
the tunneling process, the Pontrjagin index has changed from $Q=0$ to $Q=-1$.  

Within the collective coordinate approximation, the spin path integral becomes
\begin{equation}
\sexpect{{\rm FM}}{e^{-\beta H}}{{\rm FM}} \approx \int\!\CD[\lambda,\lab]\>e^{-\CAF[\lambda,\lab]}\quad,
\end{equation}
where
\begin{align}
\CAF[\lambda,\lab]&=\!\!\int\limits_0^{\beta E_0}\!\!d\tau\ \Bigg\{ S\sum_\Br {\wb_\Br\pz_\lambda w_\Br\,\pz_\tau\lambda - \,w_\Br \pz_\lab \wb_\Br \,\pz_\tau\lab  \over
1 + \wb_\Br w_\Br}\nonumber\\
&\hskip 1.4in + \Hbar(\lambda,\lab)\Bigg\}\quad,
\end{align}
where $\Hbar(\lambda,\lab)$ is obtained from $\Hbar\big(\{\BOh_\Br\}\big)$ by substituting
\begin{equation}
\sOmega_\Br^z={1-\wb_\Br w_\Br\over 1+\wb_\Br w_\Br}\quad,\quad
\sOmega_\Br^x + i\sOmega_\Br^y={2w_\Br\over 1+\wb_\Br w_\Br}\quad.
\label{Hopf}
\end{equation}
This reduced path integral over collective coordinates is dominated by the paths that extremize $\mathcal{A}_{{\rm E}}^{{\rm eff}}$, the so-called bounce instantons. Summing over all the
multi-bounce instantons \cite{Coleman1985} and the quadratic fluctuations about them, the tunneling rate from the FM to the SSk state can be approximated as
$\sGamma = C \exp\big(-\Delta\CAF\big)$, where $C$ is the fluctuation determinant prefactor, and the reduced effective action,
\begin{equation}
\Delta\CAF =  S\!\!\int\limits_0^{\beta E_0}\!\!d\tau\sum_\Br {\wb_\Br\pz_\lambda w_\Br\,\pz_\tau\lambda - \,w_\Br \pz_\lab \wb_\Br \,\pz_\tau\lab  \over 1 + \wb_\Br w_\Br}\ ,
%\label{MFeqns}
\end{equation}
is evaluated in the single-bounce instanton; here we focus on the computation of $\Delta\CAF$.  The calculation of the tunneling rate has now
been reduced to solving the Euler-Lagrange (EL) equations of motion
\begin{equation}
{d\lambda\over d\tau}=-{1\over M} {\pz\Hbar\over\pz\lab\nd} \quad,\quad
{d\lab\over d\tau} = +{1\over M}{\pz\Hbar\over\pz\lambda\nd}\ ,
\label{MFeqns}
\end{equation}
where $\Hbar$ and $M$ are functions of both $\lambda$ and $\lab$, with
\begin{equation}
M(\lambda, \bar{\lambda}) = 2S \sum_{\Br} {1\over (1 + w_{\Br}\bar{w}_{\Br})^2}\,{\pz w_\Br\over\pz\lambda\nd}\,{\pz\wb_\Br\over\pz\lab\nd}\ .
\end{equation}
Owing to the relative minus sign in Eqns. \eqref{MFeqns}, the EL equations for $\lambda(\tau)$ and $\lab(\tau)$ are not complex conjugates of each other.
Nor, since the EL equations are first order in time, are we permitted to impose boundary conditions on both $\lambda$ and $\lab$ at $\tau=0$ and $\tau=\beta E_0$.
Rather, $\lambda(\tau)$ is to be evaluated forward from initial data $\lambda(0)=\lambda_0$ (with $\lambda_0=0$ in our case, corresponding to the metastable FM state)
and $\lab(\tau)$ is to be evaluated backward from final data $\lab(\beta E_0)=\lab^*_0$, where star denotes complex conjugation.  Thus, during the bounce path, {\it $\lambda(\tau)$
and $\lab(\tau)$ are generally not complex conjugates\/}, and thus {\it the components $\sOmega^\alpha_\Br$ of local spin field, obtained in Eqn. \eqref{Hopf} from $(w_\Br,\wb_\Br)$,
are not always real\cite{PhysRevA.36.3467,PhysRevLett.79.5054}\/}.  Indeed the bounce instanton equations imply $\lab(\tau)=\lambda^*(\beta E_0-\tau)$, hence
$\wb\nd_\Br(\tau)=w^*_\Br(\tau)$ only on three time slices: $\tau=0$ and $\tau=\beta E_0$, corresponding to the FM state, and $\tau=\half\beta E_0$, where the field emerges from
the barrier and the topological charge has been nucleated.  In fact, for finite $\beta E_0$, there are exponentially small differences between $\lambda(\beta E_0)$ and $\lambda_0$,
and between $\lab(0)$ and $\lambda_0^*$.  These differences can be made arbitrarily small by increasing the value of $\beta E_0$.

\begin{figure}[!t]
\centering
\includegraphics[width=\columnwidth]{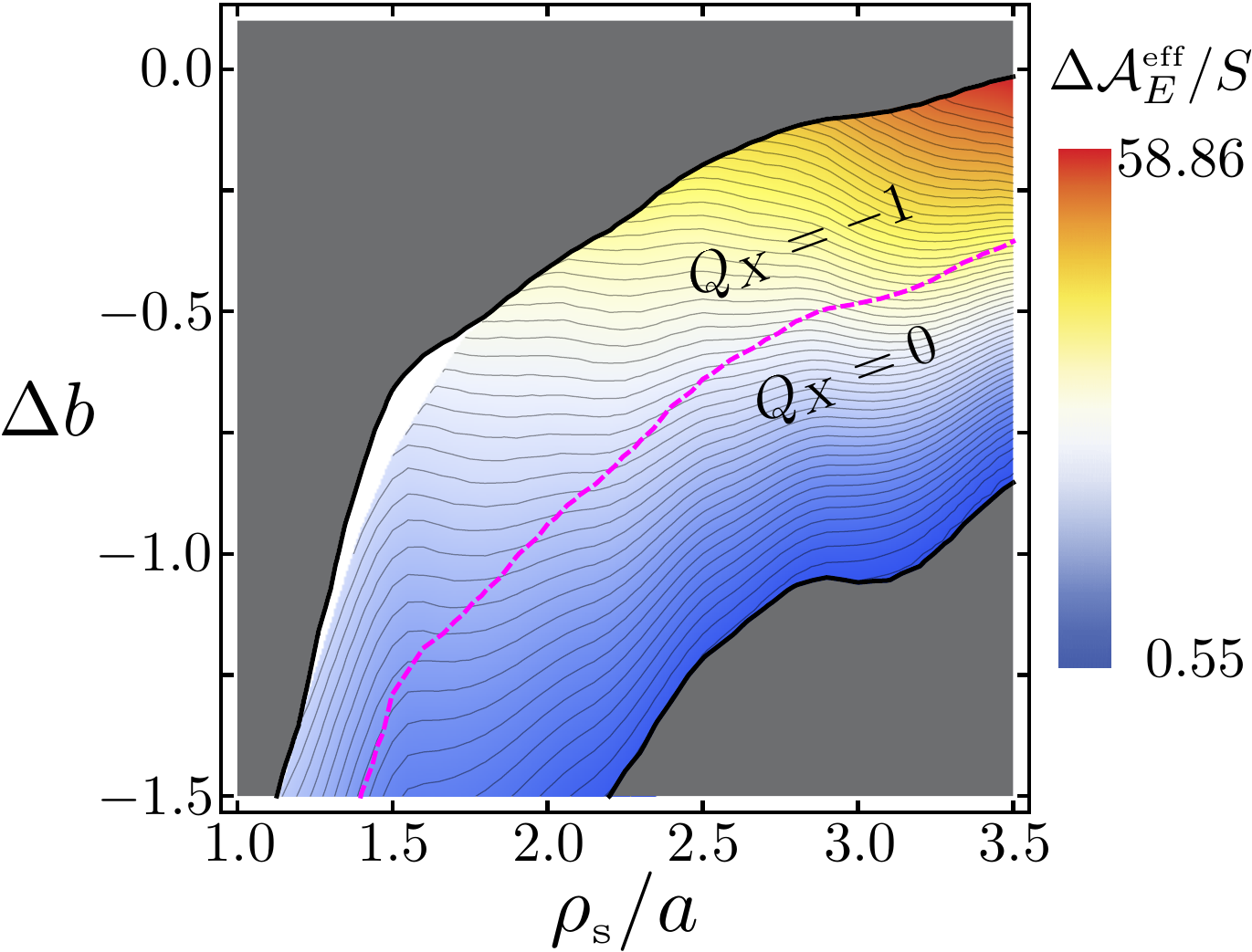}
\caption{{\bf Reduced Effective Euclidean Action}. Computed as a function of the localized magnetic field parameters $\Delta b$ and $\rho_{{\rm s}}$, for
$\{ \alpha, \kappa, b_\infty, \sigma/a \}= \{0.5, 0.1, 3.0, 0.5\}$. $S$ is the spin of the individual magnetic moments in the sample. The dashed, magenta curve separates the regions where the
X states---spin configuration quantum tunneled to from the FM state--- have $0$ and $-1$ topological charge.}
\label{fig:EffA}
\end{figure}

For a continuous family of magnetization fields $\BM(\Br,u)$ parameterized by a real number $u$, a topology change between
different Pontrjagin number sectors is possible only via Bloch points \cite{PhysRevB.88.174402,PhysRevLett.97.177202,PhysRevB.67.094410}, which are configurations where $\BM(\Br,u)$ vanishes at some location $\Br$.  This occurs for a critical value
of $u$, since this state of affairs is nongeneric, and thereby corresponds to a three-dimensional singularity such as a
hedgehog. In our tunneling formalism, the fields $w_\Br(\tau)$ and ${\bar w}_\Br(\tau)$ are in general {\it not\/} complex
conjugates of each other, and thus there is no corresponding field $\BM(\Br,\tau)$ which has a classical interpretation except
at the initial and final (imaginary) times, and at the midpoint where the fields emerge from the tunneling barrier. 
Nevertheless, if one defines the fields $\sOmega^z_\Br\equiv (1-w^*_\Br w\ns_\Br)/(1+w^*_\Br w\ns_\Br)$, $\sOmega^+_\Br\equiv
2w_\Br/(1+w^*_\Br w\ns_\Br)$, ${\bar\sOmega}^z_\Br\equiv (1-{\bar w}\ns_\Br {\bar w}^*_\Br)/
(1+{\bar w}\ns_\Br {\bar w}^*_\Br)$, and  ${\bar\sOmega}^+_\Br\equiv 2{\bar w}^*_\Br/
(1+{\bar w}\ns_\Br {\bar w}^*_\Br)$, where bar {\it does not\/} signify complex conjugation, one has that $\BOh(\Br,\tau)$
and ${\bar\BOh}(\Br,\tau)$ {\it each\/} go though Bloch points at different times, in the continuum limit.  Again, the fact that
our model is defined on a lattice avoids any actual singularities.

%%%%%%%%%%%%%%%%%%%%%%%%%%%%%%%%%%%%%%%%
%%%%%%%%%%%%%%%%%%%%%%%%%%%%%%%%%%%%%%%%

\begin{figure}[!t]
\centering
\includegraphics[width=\columnwidth]{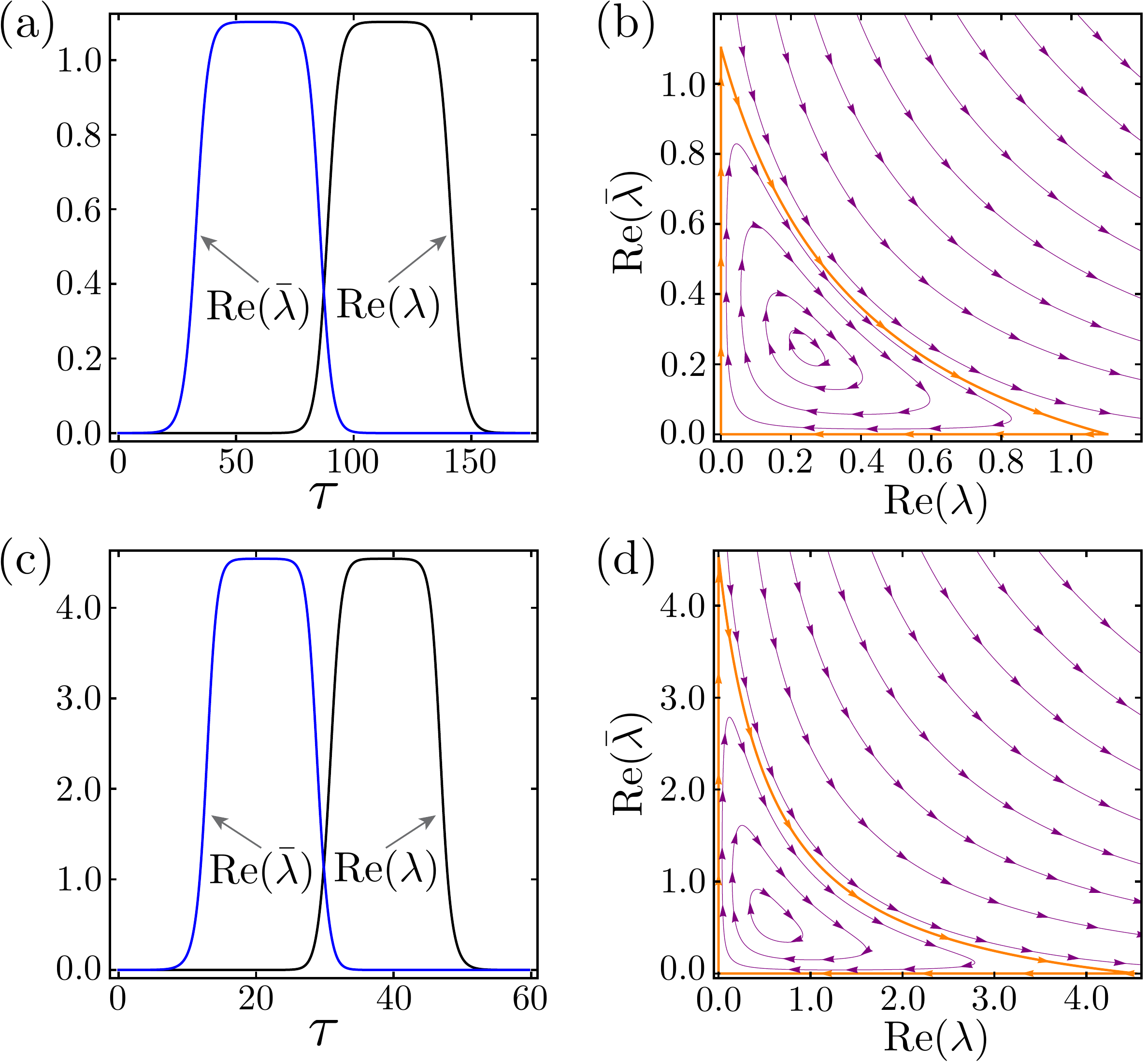}
\caption{{\bf Bounce Instantons}. (a) \& (c), bounce instantons as a function of imaginary time. Only the real part of $\lambda$ and $\bar{\lambda}$ are plotted, their imaginary parts vanished within numerical error. (b) \& (d), phase portraits of the instanton ODEs, Eqns. \eqref{MFeqns}; the orange contours correspond to the bounce instantons. Parameters $\{ \rho_{\rm s}/a, \Delta b \} = \{ 2.5, -1.0 \}$ and  $\{ \rho_{\rm s}/a, \Delta b \} = \{ 2.0, -0.8 \}$ were used for (a) \& (b) and (c) \& (d), respectively. For all panels $\{ \alpha, \kappa, b_\infty, \sigma/a \}= \{0.5, 0.1, 3.0, 0.5\}$, and $\tau$ is measured in units of $\hbar S/E_0$.}
\label{fig:PhPort}
\end{figure}

\section{Results and Discussion}

The reduced effective Euclidean action was computed as a function of the localized magnetic field parameters $\Delta b$ and $\rho_{{\rm s}}$ for the same model parameters used in the
construction of the phase diagram. Our results are presented as a contour plot in Fig. \ref{fig:EffA}. Since $\sGamma \propto \exp\big(-\Delta\CAF\big)$, a large/small $\Delta\CAF$ corresponds to
a small/large tunneling rate. As mentioned above, $\Delta\CAF$ is simply the sum of the (complex) Berry phases accumulated by all the spins in the lattice during a single-bounce instanton.
The spin Berry phase has the geometrical interpretation of the solid angle swept by the spin as it evolves in time. Therefore, $\sGamma$ is strongly dependent on the effective number of spins
flipped during the tunneling process. Further confirmation comes from analyzing the dependence of the tunneling rate on $\rho_{{\rm s}}$. The observed oscillatory trend of the $\Delta\CAF$ contours is a result, as was the case with the phase diagram (see Fig. \ref{fig:PhDiag}), of the localized magnetic field resolving the lattice sites
as $\rho_{{\rm s}}$ increases. In the vicinity of the center of the SSk, the spins tend to order opposing the external magnetic field. Consequently, as expected from the solid angle interpretation
of $\Delta\CAF$, increasing $\Delta b$ decreases the tunneling rate because it favors the alignment of spins in the direction of the external magnetic field. When the localized magnetic field
parameters approach the FM metastable-to-unstable boundary, the tunneling rate increases because the energy barrier height disappears at said boundary.

The dashed, magenta curve in Fig. \ref{fig:EffA} splits the $\Delta\CAF$ plot into two topologically distinct regions of the $X$ state, where the system has emerged from under the
tunneling barrier and where the configuration $\{\BOh_\Br\}$ has a classical interpretation (see Fig. \ref{fig:QT}).  The topological charge of this configuration, $Q_X$, can be either $0$
or $-1$. In the $Q_X = -1$ part of the plot, the FM state tunnels to a state that is already a SSk, but not the one that minimizes the energy. The minimum energy state is then
reached via classical evolution which deforms the magnetic texture, but preserves its topological charge of $-1$. On the other hand, within the $Q_X = 0$ region, the X state is not a SSk.
Interestingly, the system will evolve classically changing its topological charge to $-1$ at some intermediate point as it rolls toward the minimum energy SSk configuration. Bounce instantons representative of each of these two regions have been plotted in Fig. \ref{fig:PhPort} along with the respective phase portrait of solutions determined by Eqns. \eqref{MFeqns}.    

%%%%%%%%%%%%%%%%%%%%%%%%%%%%%%%%%%%%%%%%
%%%%%%%%%%%%%%%%%%%%%%%%%%%%%%%%%%%%%%%%
\section{Conclusions}

Quantum nucleation of individual skyrmions in magnetic ultra-thin films with interfacial DMI was studied. At zero temperature, a localized magnetic field applied to a sample in the FM state can render a SSk energetically more favorable. The magnitude of the localized magnetic field and the surface of the sample exposed to it control the size of the SSk to be nucleated. Using spin path integrals and a collective coordinate approximation, the tunneling rate from the metastable FM state to the SSk state was computed as a function of the localized magnetic field parameters. Nucleating a SSk from the FM state is unavoidably accompanied by the creation of topological charge, a process forbidden in the continuum. However, since the magnetic state of the system was described by spins on a lattice, we are able to model the quantum nucleation process using the continuous evolution of the spins. Model parameters leading to tunneling rate values that could result in observable skyrmion nucleation events were determined.

%%%%%%%%%%%%%%%%%%%%%%%%%%%%%%%%%%%%%%%%
%%%%%%%%%%%%%%%%%%%%%%%%%%%%%%%%%%%%%%%%
\appendix

%%%%%%%%%%%%%%%%%%%%%%%%%%%%%%%%%%%%%%%%
%%%%%%%%%%%%%%%%%%%%%%%%%%%%%%%%%%%%%%%%

\section{Single Skyrmion State}\label{appendix}
The continuum limit of our dimensionless energy density (units of $E_0/ta^2$) is given by
\begin{align}
\CE(\Br)&=\half (\Bnabla\BOh)^2 + \alpha\big[\sOmega^z\Bnabla\cdot\BOh - \BOh\cdot\Bnabla\sOmega^z\big]\\
&\hskip 1.4in  - \kappa (\sOmega^z)^2 - b(\Br)\,\sOmega^z\ .\nonumber
\end{align}
The single skyrmion state that extremizes the above magnetic energy functional is determined by postulating the following axially symmetric magnetization field direction
\begin{equation}
\BOh(\Br) = \sin\theta(\rho)\bs{\hat{\rho}}(\varphi + \gamma) + \cos\theta(\rho)\zhat,
\end{equation}
where $\Br = \rho\bs{\hat{\rho}}$ and $\bs{\hat{\rho}}(\varphi) = \xhat\cos\varphi + \yhat\sin\varphi$. Here $\rho$ and $\varphi$ are the usual cylindrical coordinates defined in the $xy$ plane. The helicity $\gamma$ and the skyrmion radial profile, $\theta(\rho)$, are derived from the Euler-Lagrange equations. It is found that $\gamma = 0$, which corresponds to a hedgehog-like skyrmion, while $\theta(\rho)$ is the solution to the ODE
\begin{equation}\label{eq:AdimProfEq}
\frac{d^2\theta}{d\rho^2} + \frac{1}{\rho}\frac{d\theta}{d\rho} + \frac{2\alpha\sin^2\theta}{\rho} - \left( \frac{1}{2\rho^2} + \kappa \right)\sin2\theta - b(\rho)\sin\theta = 0,
\end{equation}
with boundary conditions $\theta(0) = \pi$ and $\theta(\infty) = 0$. This ODE was solved numerically using the shooting method. The topological charge of this texture is $Q = -1$.

%%%%%%%%%%%%%%%%%%%%%%%%%%%%%%%%%%%%%%%%
%%%%%%%%%%%%%%%%%%%%%%%%%%%%%%%%%%%%%%%%

\begin{acknowledgments}
SD acknowledges partial support from the International Fulbright Science and Technology Award. 
DPA is grateful for support from the UCSD Academic Senate.  We are grateful to O. Tchernyshyov for helpful comments.
\end{acknowledgments}

%%%%%%%%%%%%%%%%%%%%%%%%%%%%%%%%%%%%%%%%
%%%%%%%%%%%%%%%%%%%%%%%%%%%%%%%%%%%%%%%%

\bibliography{SN_final}

\end{document}